\documentclass[aps,prb,twocolumn,floatfix,showpacs]{revtex4}

\usepackage{graphicx}% Include figure files
\usepackage{dcolumn}% Align table columns on decimal point
\usepackage{bm}% bold math
\usepackage{verbatim}

\newcommand{\Avec}[1]{\ensuremath{\mathbf{#1}}}

\newcommand{\Vac}[2]{\ensuremath{{#1}\text{V}_{\text{#2}}}}

\newlength{\figwidth}

\begin{document}

\title{Ground-state structure of the hydrogen double vacancy on Pd(111)}

\author{Sungho~Kim} 
\author{Seong-Gon~Kim} \email{kimsg@ccs.msstate.edu} 
\affiliation{
  Department of Physics and Astronomy, 
  Mississippi State University,
  Mississippi State, MS 39762, USA
}
\affiliation{%
  Center for Advanced Vehicular Systems, 
  Mississippi State University, 
  Mississippi State, MS 39762, USA
}

\author{S.~C.~Erwin}
\affiliation{%
  Center for Computational Materials Science, 
  Naval Research Laboratory, 
  Washington, DC 20375, USA
}

\date{June 26, 2007}

\begin{abstract}
  
  We determine the ground-state structure of a double vacancy in a
  hydrogen monolayer on the Pd(111) surface.  We represent the double
  vacancy as a triple vacancy containing one additional hydrogen atom.
  The potential-energy surface for a hydrogen atom moving in the
  triple vacancy is obtained by density-functional theory, and the
  wave function of the fully quantum hydrogen atom is obtained by
  solving the Schr\"odinger equation.  We find that an H atom in a
  divacancy defect experiences significant quantum effects, and that
  the ground-state wave function is centered at the hcp site
  rather than the fcc site normally occupied by H atoms on
  Pd(111). Our results agree well with scanning tunneling microscopy
  images.

\end{abstract}


\maketitle

\section{Introduction}

Understanding the fundamental mechanisms that govern the behavior of
hydrogen on metal surfaces is important to new technologies involving
hydrogen production, storage, and energy conversion\cite{Huber:2003,
  Schlapbach:2001, Steele:2001, Chalk:2000}.  Because of its light
mass, hydrogen can manifest uniquely quantum effects not seen for
other elements.  These effects can have striking consequences when the
dimensionality of the allowed motion is constrained by adsorption on a
surface\cite{Lauhon:2000, Cao:1997, Zhu:1992, Kallen:2001, Gross:1996,
  Chen:1994}.  When a metal surface such as palladium is covered by a
nearly complete monolayer of hydrogen, small clusters of vacancies in
the hydrogen layer serve as active sites for the dissociative
adsorption of additional H$_2$ molecules\cite{Conrad:1974,
  Holloway:2003}.  Despite their central role in catalysis, the exact
nature of these vacancy defects is not well understood.

It has been generally accepted that the dissociative adsorption of the
diatomic molecule H$_2$ is to require at least two adjacent and empty
atomic adsorption sites (or vacancies) based on early studies of
Langmuir\cite{Conrad:1974, Langmuir:1916}.  Recently, intriguing
observations about hydrogen divacancies on palladium were made using
scanning tunneling microscopy (STM).  Mitsui and co-workers
\cite{Mitsui:2003a, Mitsui:2003b} reported that hydrogen molecules
impinging on an almost H-saturated Pd(111) surface did not adsorb in
two-vacancy sites.  Rather, aggregates of three or more
hydrogen vacancies were required for efficient dissociation of H$_2$
molecules.  These findings led to speculations that the standard
description from Langmuir adsorption kinetics \cite{Conrad:1974} might
be too simple to explain the dissociative adsorption of hydrogen on
Pd(111) \cite{Mitsui:2003a, Mitsui:2003b, Holloway:2003, Lopez:2004}.

However, Gro\ss\ and Dianat have recently shown, using \textit{ab
  initio} molecular dynamics (AIMD) simulations, that no change in the
original Langmuir picture is warranted \cite{Gross:2003}.  Based on
more than 4000 AIMD trajectories, Gro\ss\ \textit{et al.}
unambiguously demonstrated that a divacancy is still sufficient to
dissociate hydrogen molecules provided the kinetic energy of the
incident molecules is large enough to overcome the relatively small
energy barrier.  For an initial kinetic energy of 0.02~eV, roughly
corresponding to the kinetic energy of the H$_2$ molecules in Mitsui's
experiment, no dissociative adsorption events were observed on the
almost hydrogen-covered Pd(111) surface, thus confirming the
experimental findings.  When the initial kinetic energy was increased
to 0.1~eV so that the small barrier found for H$_2$ on Pd(111) at high
coverage \cite{Lopez:2004} can be overcome, a nonvanishing adsorption
probability was observed.

To further improve our understanding of dissociative adsorption of
H$_2$ on Pd(111), an accurate description of the ground-state
structure of the double vacancy in a hydrogen monolayer on the surface
is desirable.  Mitsui \textit{et al.}  \cite{Mitsui:2003a,
  Mitsui:2003b} found that divacancies have a triangular STM image
extending over three neighboring sites, instead of a linear image over
two neighboring sites.  In this paper, by determining the ground-state
structure of the divacancy in a hydrogen monolayer on Pd(111), we show
theoretically that the quantum nature of the H atom accounts for the
observed STM imagery.  Our findings indicate that the quantum wave
nature of hydrogen motion provides a simple and elegant foundation to
the traditional classical explanations for such
observations\cite{Lopez:2004, Holloway:2003, Hammer:2001}, and suggest
that the quantum nature of hydrogen may play a surprisingly prominent
role in other similarly confined systems.

\section{Methods}

\begin{figure}
  \centering
  \setlength{\figwidth}{0.9\hsize}
  \begin{tabular}{cc}
    \begin{minipage}{0.5\figwidth}
      \centering
      \resizebox{\hsize}{!}{\includegraphics{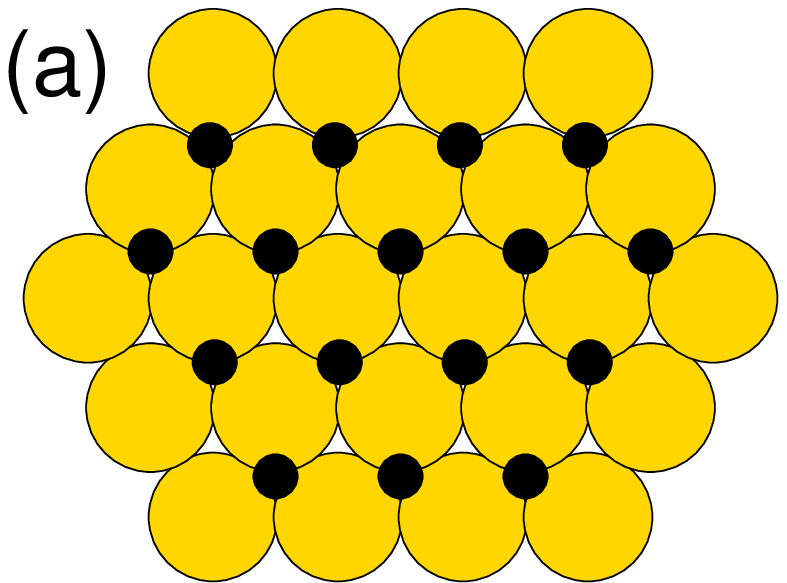}}
    \end{minipage} & 
    \begin{minipage}{0.5\figwidth}
      \centering
      \resizebox{\hsize}{!}{\includegraphics{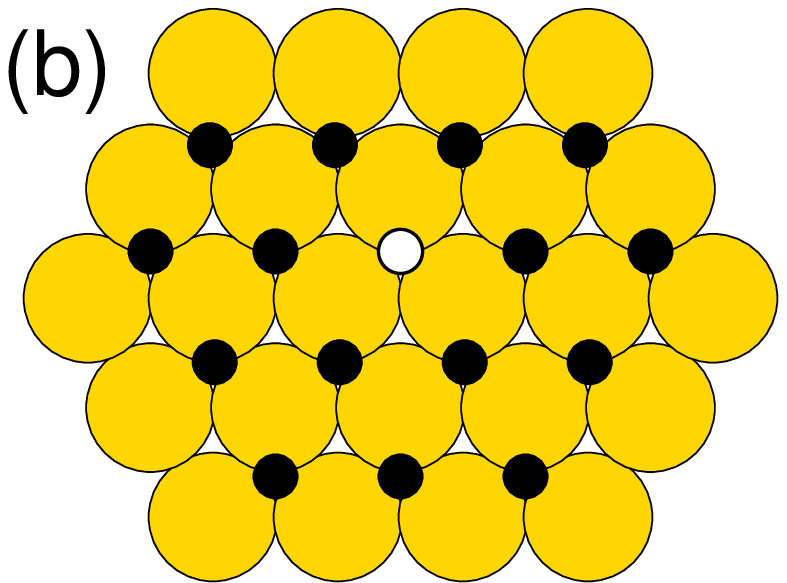}}
    \end{minipage} \\[14mm]
    \begin{minipage}{0.5\figwidth}
      \centering
      \resizebox{\hsize}{!}{\includegraphics{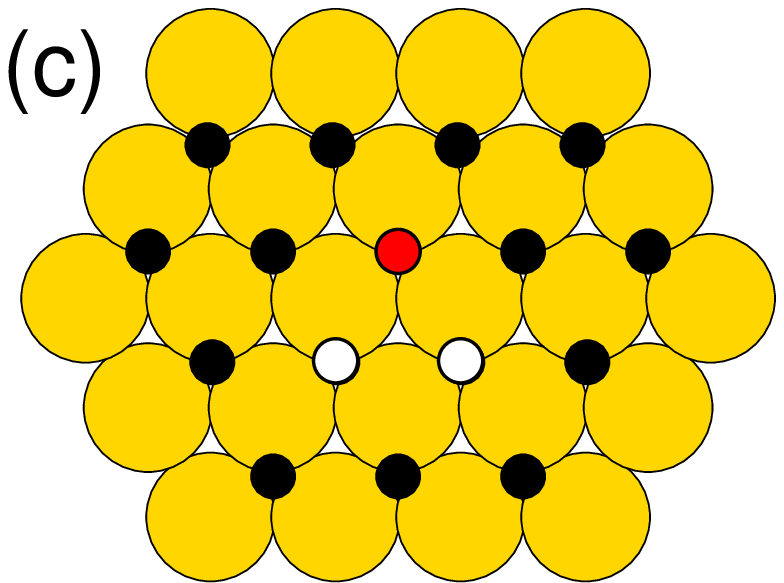}}
    \end{minipage} &
    \begin{minipage}{0.5\figwidth}
      \centering
      \resizebox{\hsize}{!}{\includegraphics{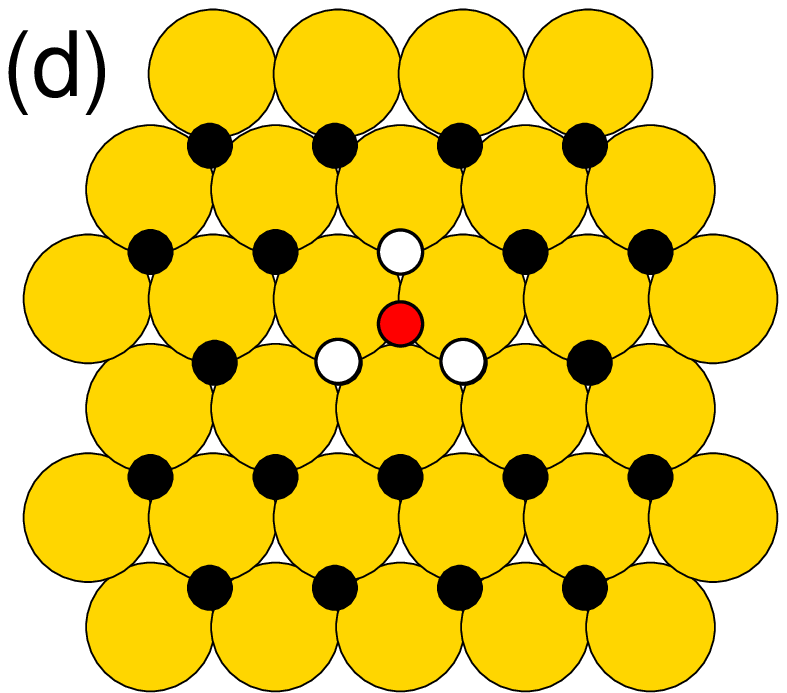}}
    \end{minipage} \\[10mm]
    \begin{minipage}{0.5\figwidth}
      \centering
      \resizebox{\hsize}{!}{\includegraphics{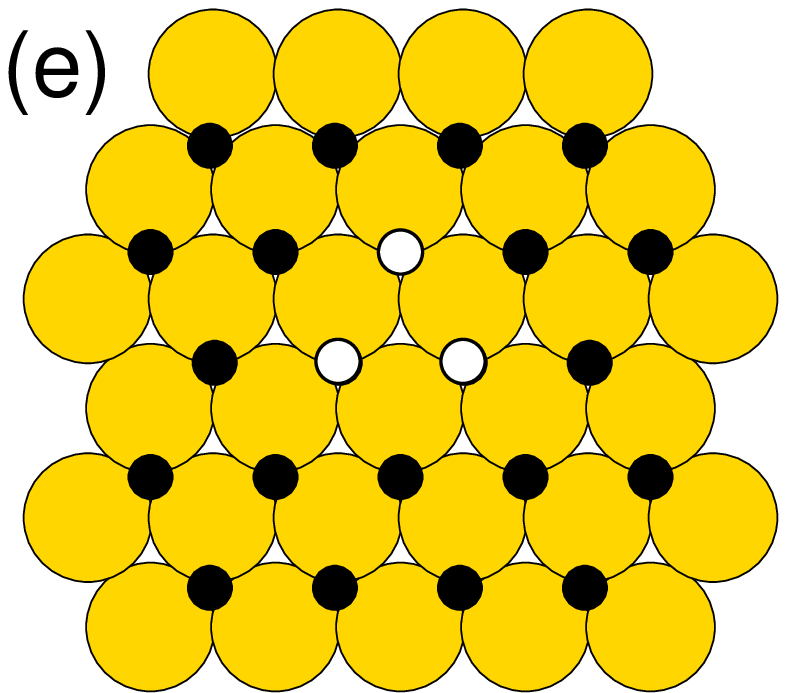}}
    \end{minipage} & 
    \begin{minipage}{0.5\figwidth}
      \centering
      \resizebox{\hsize}{!}{\includegraphics{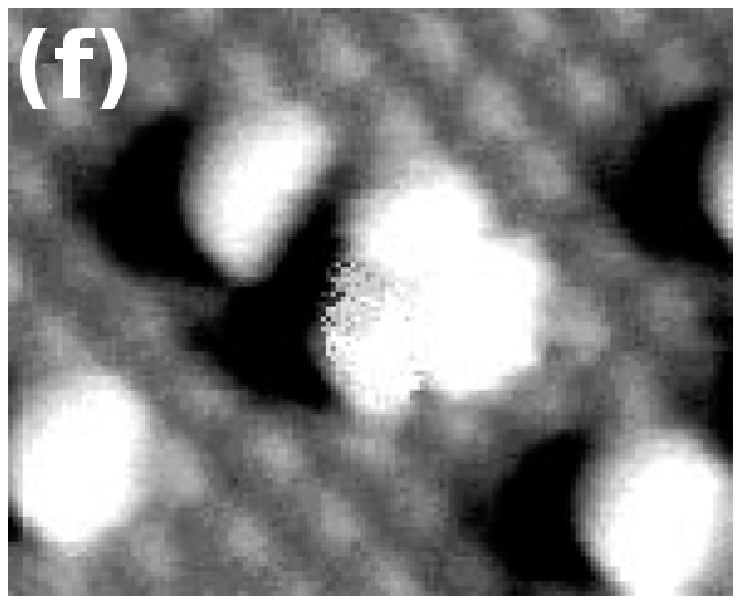}}
    \end{minipage}
  \end{tabular}
  \caption{\label{fig:H-vacancy} (color online) Hydrogen vacancy
    configurations: (a) fully hydrogenated Pd(111) surface; (b) single
    vacancy, \Vac{1}{}; (c) classical model for a divacancy,
    \Vac{2}{c}; (d) ground-state configuration of the quantum
    mechanical divacancy, \Vac{2}{q}; (e) trivacancy centered around a
    hcp site, (\Vac{3}{H}). Pd atoms are represented by large gold
    circles and H atoms by smaller black circles.  The white circles
    represent empty fcc sites while the red circle represents a H atom
    that moves in the triangular area formed by \Vac{3}{H}. (f) An STM
    image of a divacancy.  Reprinted from
    Ref.~\onlinecite{Mitsui:2003a} with permission.}
\end{figure}

Fig.~\ref{fig:H-vacancy} shows a schematic representation of several
simple hydrogen vacancy configurations on Pd(111) considered in the
present work.  A double vacancy can be considered as a single H atom
(red circles in Fig.~\ref{fig:H-vacancy}) moving in the triangular
area formed by \Vac{3}{H}.  As noted by Mitsui \textit{et al.}, this H
atom is much more mobile than all other H atoms that form
\Vac{3}{H}.\cite{Mitsui:2003a, Mitsui:2003b} Previous models that
treated H atoms classically considered only \Vac{2}{c} where the (red)
H atom hops among three fcc sites via thermal
diffusion.\cite{Mitsui:2003a} As we will show later, however, full
quantum treatment of the mobile H atom reveals that the ground-state
configuration for divacancy is \Vac{2}{q} where the central H atom
occupies the hcp site rather than fcc sites.  Exchange with the H
atoms forming the walls of the triangle of \Vac{3}{H} requires hopping
near top Pd sites and/or close to other H atoms, a process with a
higher barrier and therefore much slower than the motion of the H atom
\textit{in} the vacancy.  Hence, the problem of determining the
minimum energy configuration of a divacancy is equivalent to solving
the Schr\"odinger equation for a single hydrogen atom moving in the
external potential of hydrogen and palladium atoms having the
\Vac{3}{H} configuration.

Within the Born-Oppenheimer approximation\cite{Levine:2000}, the
Schr\"odinger equation for the motion of a hydrogen atom with mass
$M$ is
\begin{equation}
  \label{eq:SEQ}
  -\frac{\hbar^2}{2M} \nabla^2\psi(\Avec{r}) + 
  U(\Avec{r})\psi(\Avec{r}) = E\psi(\Avec{r}).
\end{equation}
The 3-dimensional potential energy surface (PES), $U(\Avec{r})$,
contains the contributions to the total energy from all Pd atoms in
the slab, all hydrogen atoms adsorbed on the surface, and the hydrogen
atom in question at $\Avec{r}$.  When only the position of this
hydrogen atom is varied, $U(\Avec{r})$ becomes, in effect, the
potential energy surface for its motion.

The PES was mapped out by calculating the adsorption energy of the
hydrogen atom placed at different positions over a \Vac{3}{H} defect.
We formed the defect within a periodic 3$\times$3 surface unit cell,
which is large enough to prevent significant interaction with periodic
images in neighboring cells. The potential energy surface was sampled
according to the importance of each region: sample points were densely
distributed near points of interest such as the fcc sites, the hcp
sites, and along the pathway between them, while points far from these
locations were sampled less densely.  Altogether over 3500 energy
points were evaluated, distributed within about 60 planes parallel to
the surface and separated by 0.1~\AA.

All \textit{ab initio} total-energy calculations and geometry
optimizations are performed within density functional theory (DFT)
\cite{Kresse:1996:VASP:PRB-69, Kohn:1965} using ultrasoft
pseudopotentials (USPP) \cite{Vanderbilt:1990} as implemented by
Kresse et. al.\cite{Kresse:1994:VASP:USPP, Kresse:1996:VASP:PRB-69}
Exchange-correlation effects were treated within the local-density
approximation (LDA)\cite{Ceperley:1980, Perdew:1981}.  The electron
wave functions were expanded in a plane-wave basis set with cutoff
energy 250~eV.  The resulting equilibrium lattice constant of bulk Pd
was 3.86 \AA\, in excellent agreement with the experimental value of
3.89~\AA\cite{Kittel:1986}.  We used a standard supercell technique to
model the Pd(111) surface as slabs of four Pd layers separated by 12
\AA\ of vacuum.  The surface Brillouin zone was sampled using a
6$\times$6 Monkhorst-Pack $k$-point set. Structural optimization was
performed until the energy difference between successive steps was
less than 1 meV.  After applying a well-known correction for the LDA
binding energy of H$_2$\cite{Dong:1996}, we obtained the hydrogen
adsorption energy $E_{\text{ad}}$=0.46~eV, in excellent agreement with
the experimental value of 0.45~eV\cite{Christmann:1973}.  Based on
tests performed with thicker slabs and larger vacuum gaps, we estimate
the uncertainty in this energy to be roughly 0.02~eV.  Tests using the
generalized-gradient approximation (GGA)\cite{Perdew:1996} show that
$E_{\text{ad}}$ values for different sites shift rigidly.  This would
have no effect on our main results below, because the PES depends only
on differences in adsorption energies, which are unchanged.

We solved the Schr\"odinger equation (\ref{eq:SEQ}) numerically in
momentum space. Although we are modeling a hydrogen atom moving in an
isolated \Vac{3}{H} defect, we nonetheless use periodic boundary
conditions in order to take advantage of Bloch's theorem and the
convenience of using plane waves as a basis.  To minimize 
computational requirements, we used a unit cell with the full $C_{3v}$
symmetry of the underlying lattice.

The wave functions for the hydrogen atom are expanded in a plane-wave
basis with a cutoff energy of 0.3~eV, which is equivalent to a cutoff
of roughly 550~eV for electronic band-structure calculations. Our
tests show that the eigenvalues of the hydrogen-atom wave functions
are well converged with this cutoff. Since we simulate an isolated
\Vac{3}{H} defect with a large unit cell, the Brillouin zone is very
small, and hence $k$ point sampling is not important.  The eigenvalues
for different $k$ points exhibit very little dispersion, and therefore
we use just a single $k$-point at the zone center.  Finally, the wave
functions are well confined inside the potential well, with no
significant tailing outside the boundaries of the potential well,
indicating that the 3$\times$3 surface unit cell is adequate.

\section{Results}

%\section{H atoms in a divacancy}

\begin{figure}
  \centering
  \resizebox{\hsize}{!}{\includegraphics{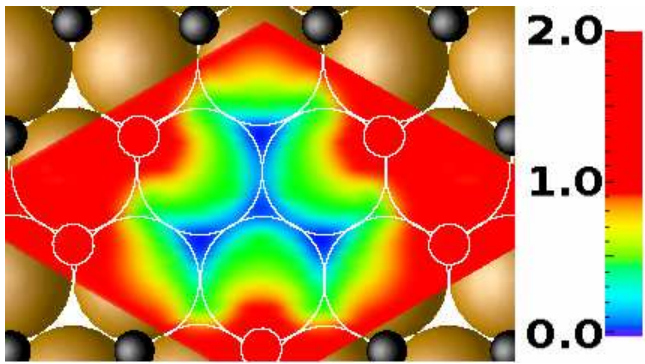}}
  \caption{\label{fig:PES} (color online) Minimum potential energy
    surface for a hydrogen atom in a divacancy. The minimum potential
    energy in the $z$-direction for all points in the $xy$-plane of a
    single classical H atom moving in a three site cluster. Large and
    small white circles representing top layer Pd atoms and
    neighboring H atoms, respectively, are drawn as visual guides.
    The color bar shows corresponding potential values in eV.}
\end{figure}

Fig.~\ref{fig:PES} shows the minimum PES for a hydrogen atom in a
\Vac{3}{H} defect.  Two important conclusions can be drawn from this
energy surface.  First, the only low-energy pathways available to the
hydrogen atom are those connecting the three fcc sites to central hcp
site.  Second, there are no low-energy pathways available to allow the
surrounding hydrogen atoms to move into the vacancy region.  The
region of low potential (blue and green) is surrounded by neighboring H
atoms and by construction possesses $C_{3v}$ symmetry about the
central hcp site. The potential floor (blue) region has three branches
extending from the central hcp site to fcc sites, where the potential
reaches its minimum.  The local minimum at the hcp site is 20~meV higher
than at the fcc sites.

\begin{table}[!tbp]
  \caption{\label{tab:eigval} The energies of eigenstates for a H atom
    in a divacancy.  The potential energies are measured from the bottom
    of the potential.  Difference in energies are measured from those
    of the ground state ($1A_1$).  States $2A_1$, $1E_1$, and $1E_2$
    are considered to be degenerate.  All energy values are given in meV.
  }
  \begin{ruledtabular}
    \begin{tabular}{ccrrrrrr}
      Index & state & $E_{\text{tot}}$ & $E_{\text{kin}}$ & $E_{\text{pot}}$ 
      & $\Delta E_{\text{tot}}$ & $\Delta E_{\text{kin}}$ 
      & $\Delta E_{\text{pot}}$ \\
      \hline
      1 & $1A_1$ & 177 & 68 & 109 & -- & -- & -- \\
      2 & $2A_1$ & 191 & 72 & 119 & 14 & 4 & 10 \\
      3 & $1E_1$ & 192 & 71 & 121 & 15 & 3 & 12 \\
      4 & $1E_2$ & 193 & 71 & 122 & 16 & 3 & 13 \\
      5 & $2E_1$ & 245 & 78 & 167 & 68 & 10 & 58 \\
    \end{tabular}
  \end{ruledtabular}
\end{table}

\begin{figure}
  \centering
  \setlength{\figwidth}{0.5\hsize}
  \begin{tabular}{cc}
    \begin{minipage}{\figwidth}
      \centering
      \resizebox{\hsize}{!}{\includegraphics{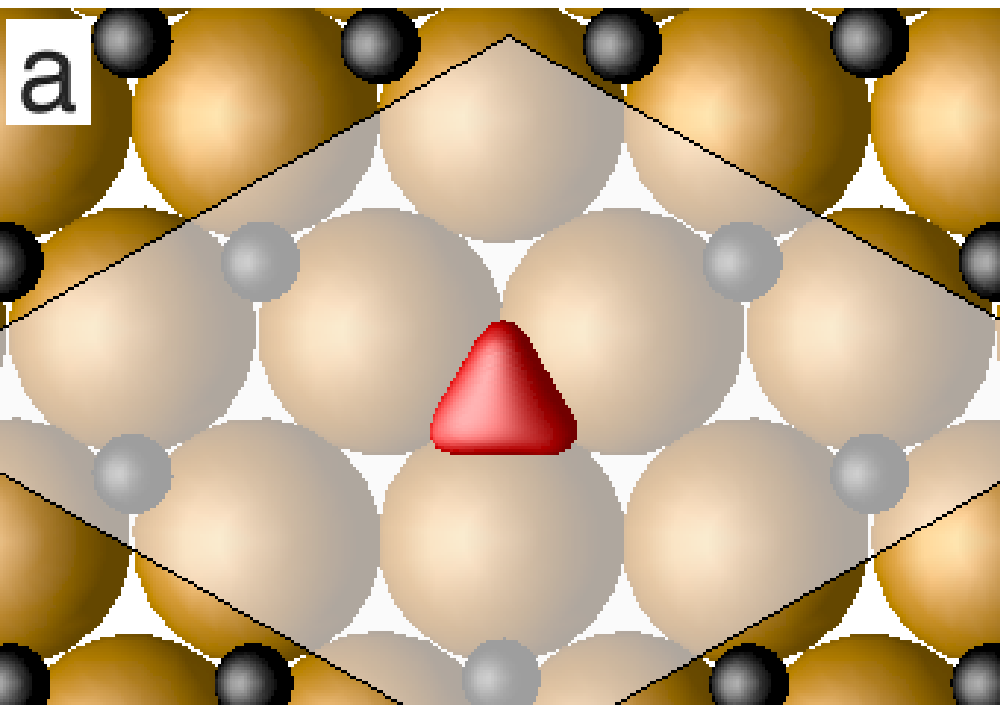}}
    \end{minipage} &
    \begin{minipage}{\figwidth}
      \centering
      \resizebox{\hsize}{!}{\includegraphics{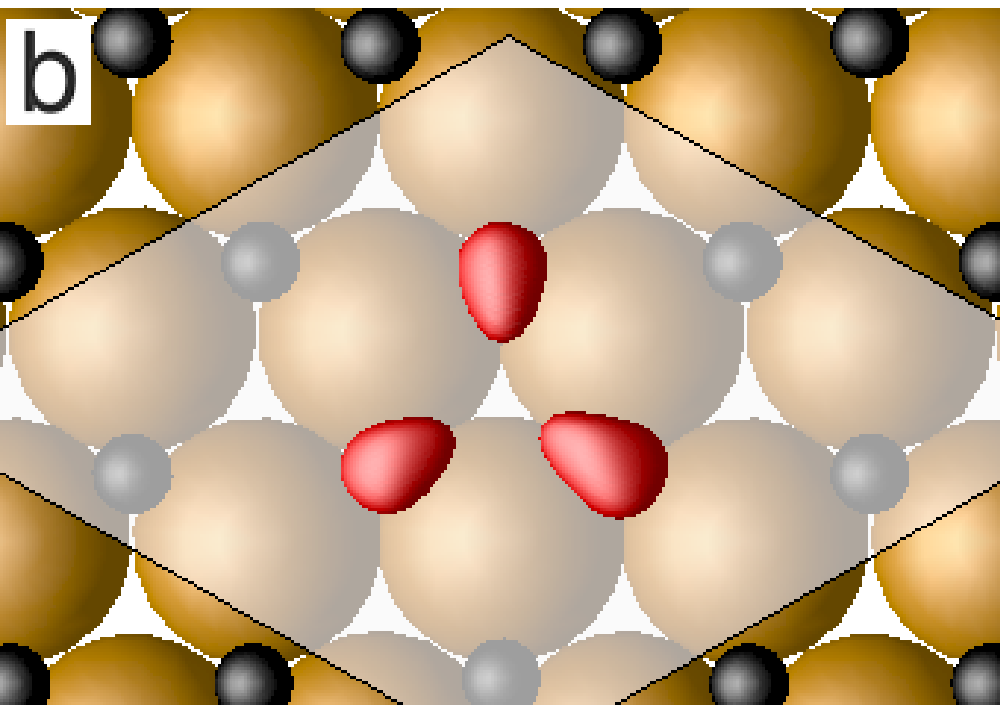}}
    \end{minipage}
  \end{tabular} \\[1mm]
  \begin{tabular}{cc}
    \begin{minipage}{\figwidth}
      \centering
      \resizebox{\hsize}{!}{\includegraphics{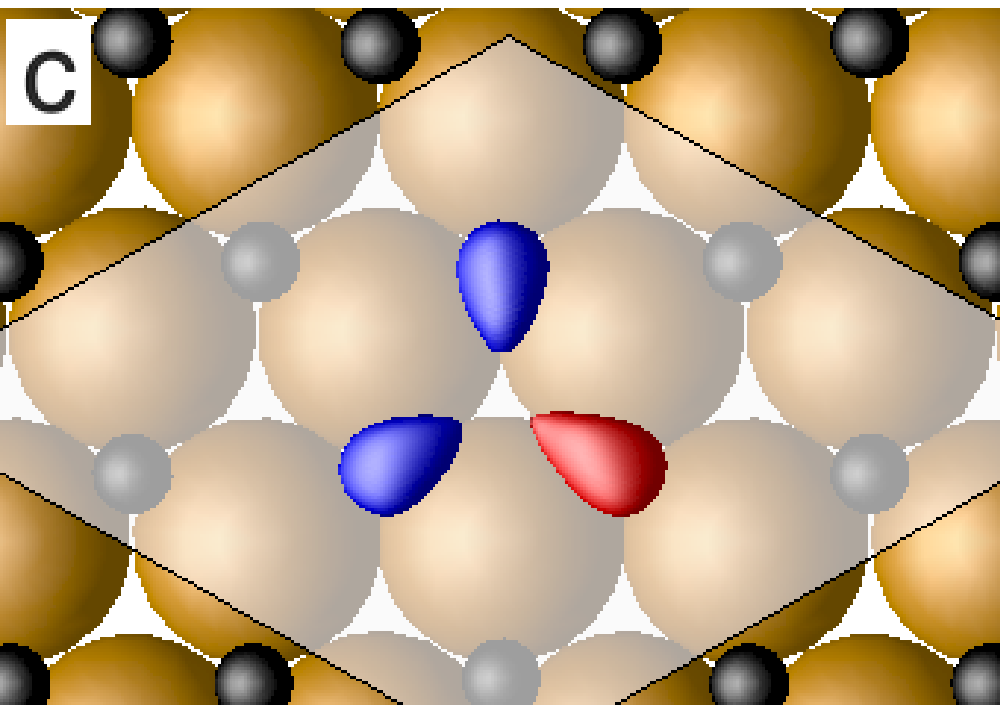}}
    \end{minipage} &
    \begin{minipage}{\figwidth}
      \centering
      \resizebox{\hsize}{!}{\includegraphics{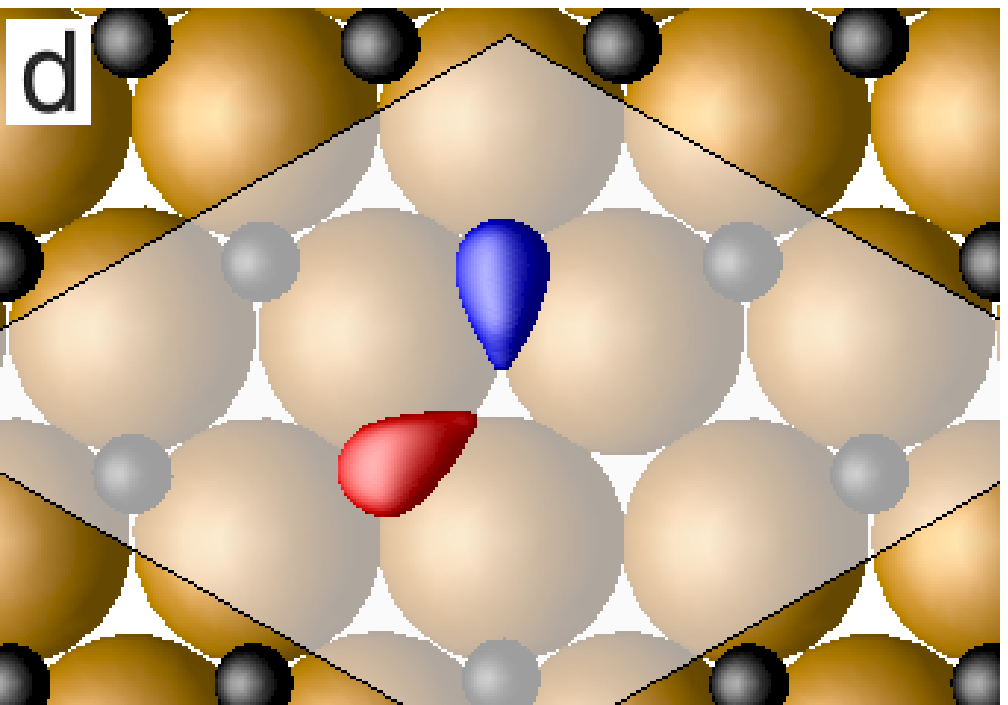}}
    \end{minipage}
  \end{tabular} \\[1mm]
  \begin{tabular}{cc}
    \begin{minipage}{\figwidth}
      \centering
      \resizebox{\hsize}{!}{\includegraphics{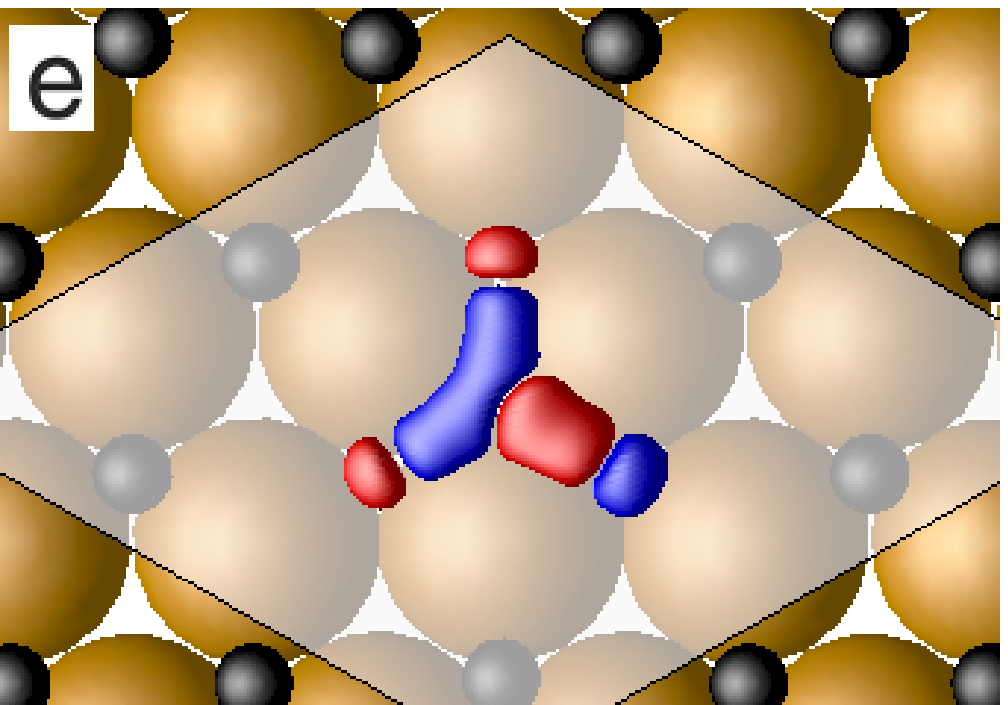}}
    \end{minipage} &
    \begin{minipage}{\figwidth}
      \centering
      \resizebox{\hsize}{!}{\includegraphics{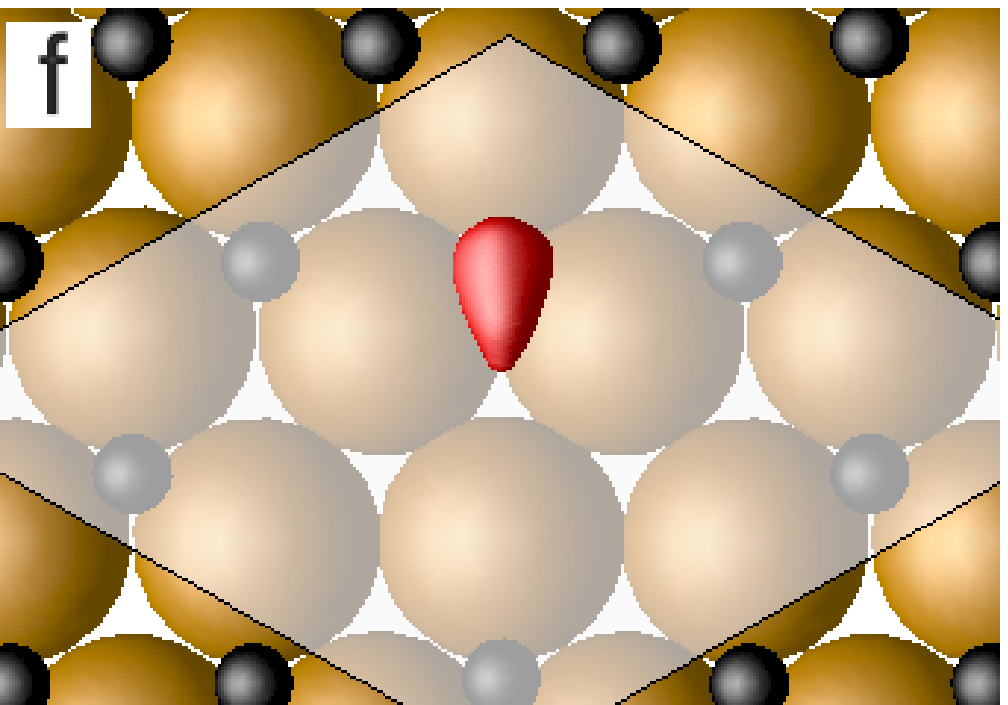}}
    \end{minipage}
  \end{tabular}
  \caption{\label{fig:2Vq} (color online) Wave functions of a H atom
    in a divacancy.  Isosurface of the real component of the wave
    functions of a hydrogen atom in a structure used to simulate a
    \Vac{2}{q} vacancy defect that contains three fcc sites and one
    hcp site. The isovalues on red surfaces are positive while the
    values on blue surfaces are negative.  All imaginary components
    are zero.  (a) The ground state, $\phi_{1A_1}$.  The red surface
    contains 99.3\% of the total integrated probability (TIP).  (b)
    The first excited state, $\phi_{2A_1}$. Three red surfaces contain
    99.1\% of the TIP.  (c) One of the first doublet excited states,
    $\phi_{1E_1}$.  The red surface contains 58.7\% of the TIP while
    two blue surfaces contain 40.3\%.  (d) The other first doublet
    excited states, $\phi_{1E_2}$.  Each surface contains 49.6\% of
    the TIP.  (e) One of the second doublet excited states,
    $\phi_{2E_1}$.  The red surface contains 61.4\% of the TIP while
    two blue surfaces contain 37.5\%.  (f) One of the mixed states
    delocalized over an fcc site: $\Psi_{\text{fcc}} =
    \frac{1}{\sqrt{6}}(\sqrt{2}\phi_{2A_1} -\phi_{1E_1}
    -\sqrt{3}\phi_{1E_2})$. The red surface contains 99.4\% of the
    TIP.}
\end{figure}

The results of our calculations for the low-energy eigenstates are
summarized in Table~\ref{tab:eigval}, and their wavefunctions are
shown in Fig.~\ref{fig:2Vq}. Of particular interest is the ground
state ($1A_1$), which is localized at the hcp site.  We note that this
is not the minimum of the potential-energy surface (which occurs at
the fcc sites).  If the H atom had been treated classically, the
ground state configuration would have the H atom adsorbed at the fcc
site, corresponding to the defect configuration \Vac{2}{c}.

To understand better the origin of this quantum effect we list
separately, in Table~\ref{tab:eigval}, the kinetic- and
potential-energy contributions to the total energy.  Our result
indicates that the quantum-mechanical shift is driven by both the
kinetic and potential energy of the hydrogen-atom wave function.  Even
though the minimum potential at the hcp site is slightly higher than
that of three fcc sites, the potential well at the hcp site is wider
than those at the fcc sites.  By localizing around the central hcp
site, the ground-state wave function can sufficiently lower its
potential energy (by reducing the penetration of wave function into
the region of higher potential) and kinetic energy (by reducing the
gradient of the wave function) to compensate the small increase of
potential around the hcp site.  Hence, the fully quantum mechanical
ground-state geometry for the hydrogen divacancy is not the classical
configuration \Vac{2}{c} of Fig.~1(c), but rather the quantum
configuration \Vac{2}{q} illustrated in Fig.~1(d).  Indeed, in order
to realize a configuration similar to the classical \Vac{2}{c}, the
quantum wave function must be localized on one of the fcc sites as
shown Fig.~\ref{fig:2Vq}(e).  This requires constructing, at the cost
of $14$~meV, a linear combination three degenerate excited states:
$\Psi_{\text{fcc}} = \frac{1}{\sqrt{6}}(\sqrt{2}\phi_{2A_1}
-\phi_{1E_1} -\sqrt{3}\phi_{1E_2})$.

\section{Discussion}

One prediction that arises from this fully
quantum-mechanical calculation is that the ground-state of the hydrogen
quantum divacancy \Vac{2}{q} will have intrinsically triangular symmetry,
suggesting that STM images should appear triangular as well. Because
it is locally insulating, an adsorbed hydrogen atom screens the
tunneling current between the STM tip and the metal substrate.  Thus,
a site occupied by an hydrogen atom has a low apparent height (dark in
a standard STM representation) while a vacancy appears high
(bright)\cite{Mitsui:2003a}.  As a result, \textit{a divacancy is
  expected to appear as a three-lobed object with triangular
  symmetry}.  Recent STM images, reproduced in
Fig.~\ref{fig:H-vacancy}(f), confirm this behavior\cite{Mitsui:2003a,
  Mitsui:2003b}. Alternative explanations based on classical thermal
diffusion\cite{Mitsui:2003a, Mitsui:2003b}, although consistent with
the observed symmetry, are not necessary to explain such images.  For
the trivacancy and larger multi-vacancy defects, the quantum effects manifested
in the divacancy are suppressed and the delocalized H atoms behave
more like classical particles. In these cases,
thermal diffusion provides a reasonable explanation for the observed STM
imagery\cite{Mitsui:2003a, Mitsui:2003b}.
  
Our results also suggest that future studies involving divacancies of
H atoms on metal (111) surfaces need to include the quantum
configuration \Vac{2}{q} shown in Fig.~\ref{fig:H-vacancy}(d) as one
of the viable stable configurations even if the total energy of the
structure when H atoms are treated classically is somewhat higher.

\section{Conclusions}

We have studied the quantum nature of hydrogen atoms on Pd(111)
surface by solving the Schr\"odinger equation for an H atom moving in
static potential energy surface determined from first-principles
density-functional theory calculations. We find that a H atom in a
divacancy defect experiences significant quantum effects, with the
result that its wave functions are extended over large portion of the
vacancy. The ground-state H wave function is
centered at an hcp site rather than the fcc site occupied by classical
H atoms. As a result, the divacancy should
exhibit a triangular geometry with three-fold symmetry, consistent
with recent experiments.

\section{Acknowledgment}

This work was in part supported by the Department of Defense
under the CHSSI MBD-04 (Molecular Packing Software for \textit{ab
  initio} Crystal Structure and Density Predictions) project and by
the Office of Naval Research.  Computer time was
provided by the High Performance Computing Collaboratory (HPC$^2$) at
Mississippi State University.

\bibliography{DFT,HPd}

\end{document}